\documentclass[10pt,conference,compsocconf]{IEEEtran}
\IEEEoverridecommandlockouts
\usepackage{subfig}
\usepackage[dvipdfmx]{graphicx}
\usepackage{latexsym}
\usepackage{nidanfloat}
\usepackage{url}
\usepackage{cite}
\usepackage{algorithm}
\usepackage{algorithmic}
\usepackage{amsmath}
\usepackage{amsfonts}

\usepackage{color}
\usepackage{multirow}

\usepackage{clrs+}

\usepackage{here}

\newcommand{\procdecl}[1]   {\proc{#1}\vrule width0pt height0pt depth 7pt \relax}

\newcommand{\New}		[1]{{{\color{black}#1}}}

\def\fig_dir{.}

\begin{document}
\begin{figure*}[t]
 \begin{center}
  \includegraphics[width=\hsize]{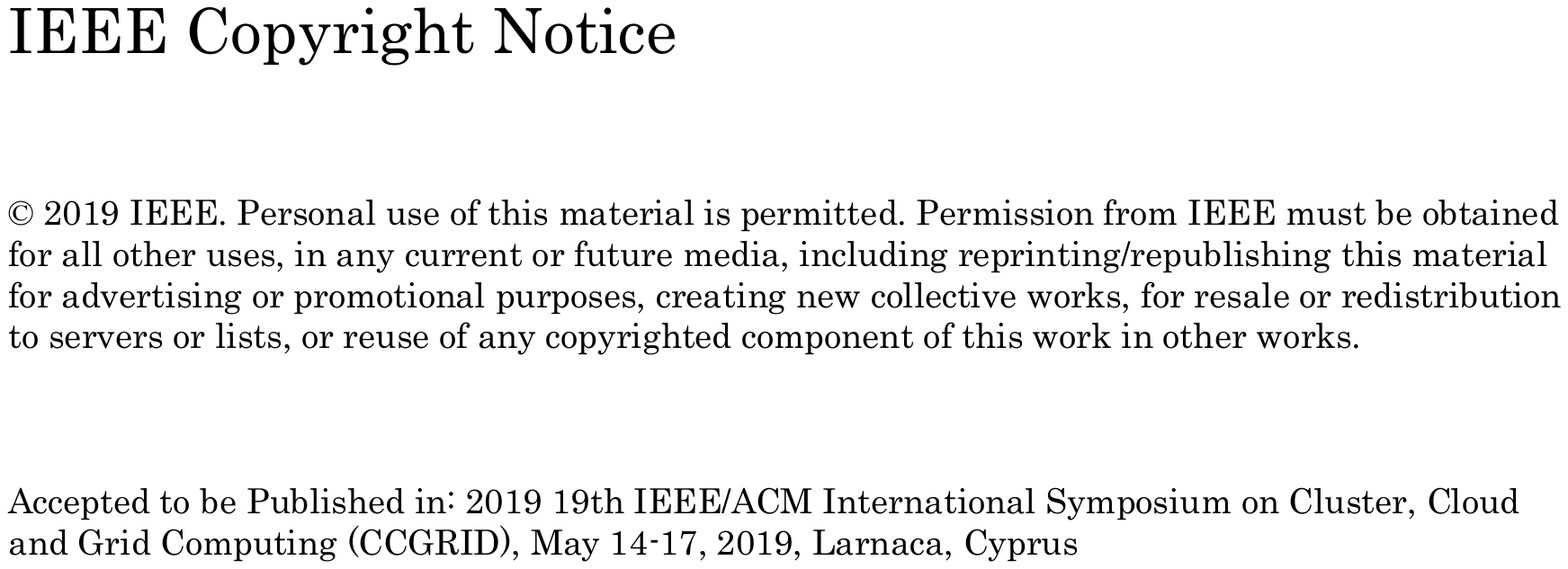}
 \end{center}
\end{figure*}

\title{Batched Sparse Matrix Multiplication for Accelerating Graph Convolutional Networks}

\author{
\IEEEauthorblockN{Yusuke Nagasaka, Akira Nukada}
\IEEEauthorblockA{
Tokyo Institute of Technology \\
Tokyo, Japan \\
nagasaka.y.aa@m.titech.ac.jp \\
nukada@gsic.titech.ac.jp}
\and
\IEEEauthorblockN{Ryosuke Kojima}
\IEEEauthorblockA{Kyoto University \\
Kyoto, Japan \\
kojima.ryosuke.8e@kyoto-u.ac.jp}
\and
\IEEEauthorblockN{Satoshi Matsuoka}
\IEEEauthorblockA{RIKEN Center for Computational Science \\
Kobe, Japan \\
matsu@acm.org}
}
\maketitle              
\begin{abstract}
Graph Convolutional Networks (GCNs) are recently getting much attention in bioinformatics and chemoinformatics as a state-of-the-art machine learning approach with high accuracy. GCNs process convolutional operations along with graph structures, and GPUs are used to process enormous operations including sparse-dense matrix multiplication (SpMM) when the graph structure is expressed as an adjacency matrix with sparse matrix format. However, the SpMM operation on small graph, where the number of nodes is tens or hundreds, hardly exploits high parallelism or compute power of GPU. Therefore, SpMM becomes a bottleneck of training and inference in GCNs applications. In order to improve the performance of GCNs applications, we propose new SpMM algorithm especially for small sparse matrix and Batched SpMM, which exploits high parallelism of GPU by processing multiple SpMM operations with single CUDA kernel. To the best of our knowledge, this is the first work of batched approach for SpMM. We evaluated the performance of the GCNs application on TSUBAME3.0 implementing NVIDIA Tesla P100 GPU, and our batched approach shows significant speedups of up to 1.59x and 1.37x in training and inference, respectively.
\end{abstract}
%
%
%

\section{Introduction}
Recently, Deep Learning is getting much attention for its high accuracy in image recognition and other machine learning topics.
In image recognition, Convolutional Neural Networks (CNNs), which consist of convolution and pooling layers, effectively expose the features of images, and show significant accuracy. On the other hand, Graph Convolutional Networks (GCNs) have been proposed~\cite{kipf2016semi} and can deal with a general graph structure, not only for a regular grid structure such as image. The GCNs show high accuracy in bioassay to predict the characteristics of chemical compounds or protein by treating the structures of substances as graph~\cite{duvenaud2015convolutional, li2015gated, kearnes2016molecular, gilmer2017neural, faber2017machine}. Furthermore, a knowledge graph is also applied to GCNs~\cite{schlichtkrull2018modeling}.

As well as CNNs, GCNs require enormous computational operations, and GPUs with high compute capability are crucial component for the GCNs applications. The structure of input graph is expressed as an adjacency matrix or adjacency list. As the connectivity between nodes among the graph is usually sparse, the adjacency matrix is also sparse.
In order to achieve high throughput in training and inference of GCNs applications, the operations on sparse matrix, especially sparse-dense matrix multiplication (SpMM), need to be accelerated. However, the dataset used in GCNs often includes very small graphs, where the number of nodes is less than one hundred. SpMM operation on such small matrix hardly exploits massive parallelism of GPU.
Furthermore, it is not clear how to efficiently process SpMM between small matrices since no research in high performance computing or other fields focuses on small sparse matrix. The researchers and developers are forced to process the training or inference of GCNs with low throughput SpMM on small matrices. As a result, the operations on sparse matrix, especially SpMM, become the bottlenecks in GCNs applications.

We propose Batched SpMM, which increases the parallelism and attains whole computing power of GPU.
Firstly, we devise new efficient SpMM algorithm, named Sub-Warp-Assigned (SWA) SpMM, for the data structure such as CSR and SparseTensor in TensorFlow, and integrate cache blocking optimization for utilizing shared memory.
Next, we show how Batched SpMM assigns computing resources.
The Batched SpMM appropriately applies cache blocking optimization and assigns threads and shared memory to each SpMM operation based on the matrix sizes in the batch.
We also mention how to efficiently apply our batched approaches to GCNs application, improving the performance of matrix multiplication and addition besides SpMM.
The Batched SpMM launches single CUDA kernel and executes tens or hundreds of SpMM operations, corresponding to mini batch size, in parallel.
The Batched SpMM enhances the parallelism and throughput of SpMM operations, and also reduces the overhead of CUDA kernel launches.

We have preliminary evaluation of the Batched SpMM on TSUBAME3.0 implementing NVIDIA Tesla P100 GPU. Our batched approaches achieve significant speedups of up to 9.27x compared to non-batched approaches, and attain improvement of up to 6.09x speedup for larger dense input matrix. We also evaluate the performance of Batched GEMM of cuBLAS by treating input sparse matrix as dense matrix, and our Batched SpMM shows superior performance to the Batched GEMM.
This preliminary evaluation exposes the effectiveness of our Batched SpMM to specific batch size and matrix size.
The Batched SpMM is integrated to the GCNs application, and accelerates the training and inference of the application. The Batched SpMM attains the speedups of up to 1.59x and 1.37x in training and inference, respectively.

\section{Background}
\subsection{Graph Convolution}
A graph convolution is a general convolutional operation, according to a graph structure, while the conventional convolutional operation assumes regular grid structure. A graph convolutional network consists of multiple graph convolution layers.
Graph structures and input features are given as input data, and the convolutional operation is executed along with its graph structure.
More specifically, the features of adjacent nodes to the target node is convoluted with filters. This operation is formulated as below when a graph is $G=\{ V, E \}$, feature vector of the node $v \in V$ is $x_v$ and a set of filter is expressed as a weight matrix, $W$.
\begin{equation}
y_i = \sum_{j \in V} a_{ij} x_j^T W
\end{equation}
The adjacency matrix is set as $a_{uu}=1$, and if an edge from $u$ to $v$ exists, $a_{vu}=1$, otherwise 0.
Executing the operations for all nodes among the graph is regarded as the multiplication between the adjacency matrix $A$, feature vectors and weight matrix as below.
\begin{equation}
Y = AXW
\end{equation}
The adjacency matrix representing a graph structure usually shows sparse property, while the feature vectors and weight matrix are expressed as dense matrix.

\subsection{Sparse Matrix Format}
\begin{figure}[t]
 \begin{center}
  \includegraphics[width=\hsize]{\fig_dir/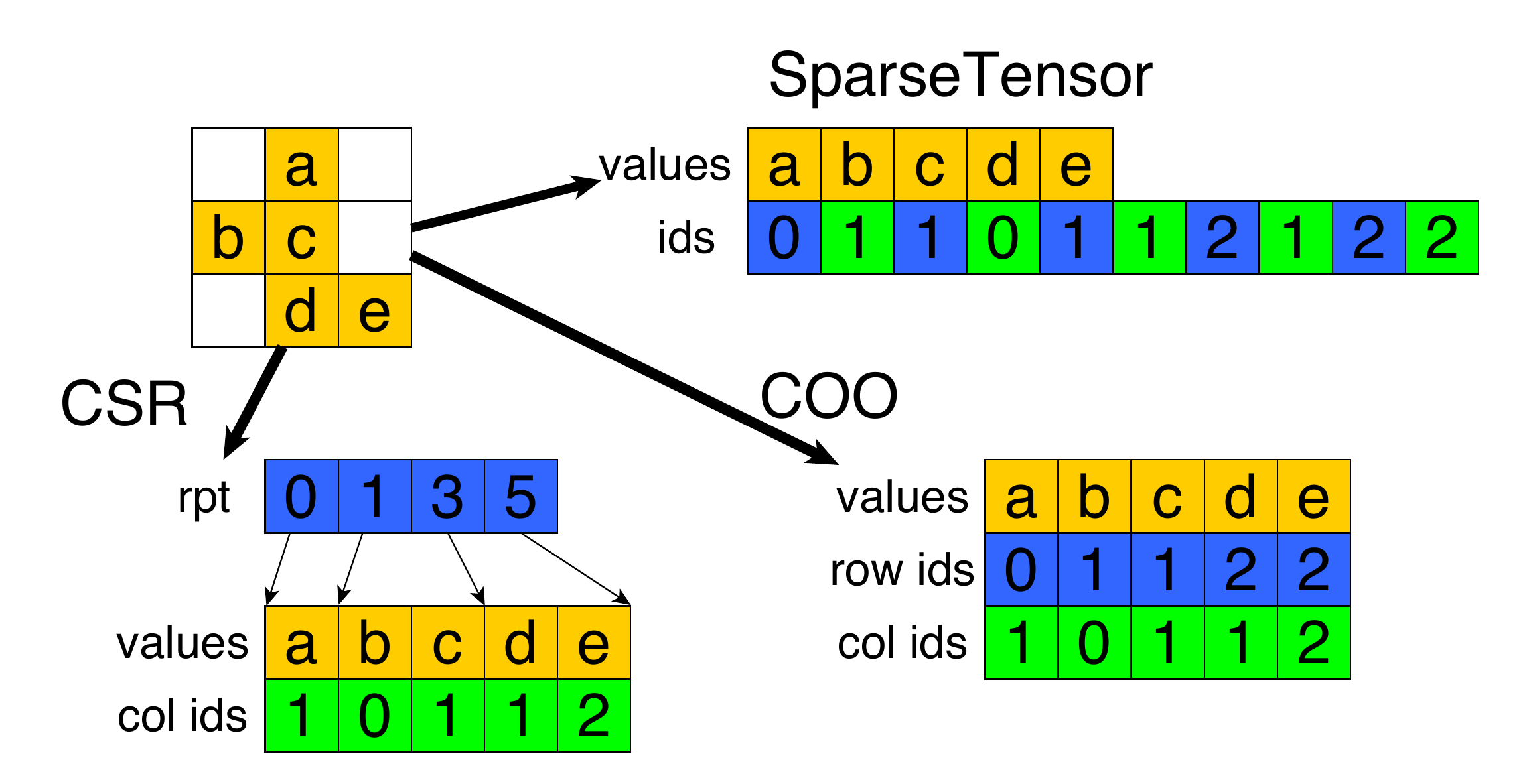}
  \caption{Example of sparse matrix formats}
  \label{fig:sparse_format}
 \end{center}
\end{figure}

Sparse matrix is usually compressed by removing zero elements and holding only non-zero elements, which are necessary for computation. Sparse matrix format is designed to reduce both operations and memory usage by compression, and various sparse matrix formats have been proposed. Coordinated (COO) and Compressed Sparse Row (CSR) are widely used.
Figure~\ref{fig:sparse_format} shows an example of each sparse matrix format. The COO format has the tuple of value, row index and column index of each non-zero element in the matrix. The CSR accumulates the non-zero elements with same row indices, and then manages the non-zero elements by holding row pointer (rpt), which indicates the beginning point of each row. The CSR can reduce memory usage compared to the COO format.
In TensorFlow~\cite{tensorflow2015-whitepaper}, one of the sophisticated deep learning frameworks, sparse matrix (tensor) is treated as SparseTensor class object. As Figure~\ref{fig:sparse_format} shows, the data structure of SparseTensor is similar to COO. In SparseTensor, the indices of each non-zero element are stored as the array of the pairs of the row and column indices.

\subsection{Sparse Matrix Multiplication}
Let A, B be input matrices, sparse matrix multiplication (SpMM) computes $C=AB$ when A is sparse matrix and B is dense matrix.
In the rest of this paper, the number of non-zero elements, row size and column size of a matrix $X$ is written in $nnz_X$, $m_X$, $n_X$, respectively.
Figure~\ref{algo:code_spmm} shows the pseudo code of SparseTensorDenseMatMul routine, which can compute SpMM in TensorFlow. The multiplications for each non-zero element are processed one after another. In CUDA code designed for GPU, two ``{\bf for}'' loops are parallelized by $nnz_A * n_B$ threads, and each thread computes a multiplication independently. 
The product is added to output matrix by using atomic operation since the memory access to output matrix and the ``add'' operation by each thread have a chance of race condition.
When the negative influence of atomic operations is enough negligible, the work assigned to each thread is about equal and the SparseTensorDenseMatMul on GPU can achieve good load balance.
However, this approach has poor locality of memory access in spite of matrix-matrix multiplication. Also, the performance is limited by the atomic operations on the global memory of GPU. To make matters worse on small matrices, the approach with $nnz_A * n_B$ threads hardly exploits enormous parallelism of GPU.
\begin{figure}[t]
  \begin{center}
\begin{codebox}
    \Procname{$\procdecl{SparseTensorDenseMatMul}(C, A, B)$}
\li \Comment set matrix $C$ to $O$
\li \For $\id{i} \gets 0$ \To $\id{nnz_A}$
\li \Do \For $\id{j} \gets 0$ \To $\id{n_B}$
\li \Do $rid \gets ids_A[i * 2]$
\li $cid \gets ids_A[i * 2 + 1]$
\li $val \gets values_A[i]$
\li $C[rid][j] \gets C[rid][j] + val * B[cid][j]$
\End
\End
\end{codebox}
  \end{center}
\vspace{10pt}
\caption{Pseudo code of the multiplication kernel between sparse tensor and dense matrix provided in TensorFlow. $A$ is stored as SparseTensor data structure.}
\label{algo:code_spmm}
\end{figure}

\section{Related work}
\subsection{SpMM for GPU}
V\'azquez and {\it et al.} proposed new approach of SpMM for GPU with ELLR-T format, which has favorable property of memory access on GPU. While the approach of SpMM with ELLR-T achieves higher performance, the format conversion from COO or CSR format causes performance degradation when the calculation on each matrix is executed only once.

Hong and {\it et al.} proposed new sparse matrix format and algorithm for SpMM named Row-Segmented-SpMM (RS-SpMM)~\cite{hong2018efficient}. The sparse matrix is divided into two groups. One is dense non-zero area and the other is rest of non-zero elements. Each part is processed by it own kernel. This approach can improve the locality of memory access and reduce the memory access to global memory. Also, in order to identify ``dense'' area, the performance model is developed and brings enough high performance even compared to best case. The format conversion including parameter setting and preprocessing for DCSR format~\cite{buluc2008representation} requires the cost equivalent to several times of SpMM calculations.

Yang and {\it et al.} proposed the state-of-the-art SpMM algorithm for GPU with CSR format~\cite{yang2018design}.
The authors proposed two kernels, and the heuristic selects between the kernels by input matrix with high accuracy. One kernel adopts merge-based approach~\cite{merrill2016merge} designed for improving the load-balance in SpMV, and the other kernel is Row-splitting, performing coalesced memory access. It is not clear that the proposed approach is effective for smaller matrices since the approach is evaluated only on large matrices.

\subsection{Batched BLAS}
The calculation on a dense matrix divided into small sub matrices~\cite{dong2014lu, haidar2015framework, charara2017batched, boukaram2018batched} or sparse matrix with block structure~\cite{zheng2015gpu, venkat2016automating, li2013exploration} is dealt with the operations on small dense matrices.
The computing kernel on small matrices suffers from the overhead of repeated kernel launches, especially on GPU, and the high parallelism and memory bandwidth of GPU are hardly exploited. Thus, the total performance of applications is decreased.
To overcome these issues, a new basic routine called Batched BLAS has been proposed handling multiple data and operations in a single kernel~\cite{abdelfattah2016performance, nath2010improved, haidar2015batched}.
Even if the data sizes in the batch are different to each, Batched BLAS keeps good load-balance. The approach for GEMM (multiplication between dense matrices) on small matrices is also proposed~\cite{masliah2016high}, and a lot of GEMM operations between small matrices can be executed with high throughput.
Recently, Batched SpMV for small matrices has been proposed\cite{anzt2017flexible}. Our batched approach is very similar concept to the Batched SpMV, but the paper about Batched SpMV includes many assumptions (eg. same non-zero pattern among matrices in batch), and the Batched SpMV is highly specialized for the application. Thus, the batched approach is very simple by adding z-axis parallelism to thread grid in CUDA, and the load imbalance among a batch is not taken into account.

\subsection{GCNs Application}
The GCNs applications are actively developed in bioinformatics or chemoinformatics areas as a practical use case of machine learning approach.
Deepchem, an open source library, is publicly available, providing the deep-learning approaches to wide range areas such as drug design and materials chemistry~\cite{deepchem}. DeepChem manages a graph structure as adjacency lists. The convolutional operations accumulate the feature vectors from adjacent nodes of the target node by exploring the adjacency lists. The convolutional operations in DeepChem hardly exploit high parallelism of GPU.

A library for deep learning in chemistry area named Chainer Chemistry is also publicly available~\cite{chainerchem}. Chainer Chemistry enables to predict the characteristics of chemical compounds by applying a molecular structure to the deep learning approaches. The data with graph structure is applied to GCNs as input data. Chainer Chemistry, however, manages an adjacency matrix as dense matrix while the matrix shows sparse property. This policy causes many redundant zero-related calculations. Furthermore, it is hard to keep the data on memory or cache when the graph becomes large.

\section{Proposal}
The GCNs application executes convolutional operations in accordance with graph structure by computing SpMM when the graph structure is expressed as sparse matrix. Learning phase of GCNs requires many SpMM executions since the GCNs need to process a lot of data through the graph convolution layers. However, SpMM on very small matrices such that the dimension is tens or hundreds hardly utilizes the high computing capability of GPU. Therefore, the overhead of CUDA kernel launch for each SpMM operation appears to dominate the total GCNs performance.
We propose Batched SpMM, which is high-throughput batched approach for processing many SpMM operations especially on small sparse matrices.
Firstly, we devise new efficient SpMM algorithm, named Sub-Warp-Assigned (SWA) SpMM, for the data structures such as CSR and SparseTensor in TensorFlow. Next, we show the cache blocking optimization to the SWA SpMM for utilizing fast shared memory.
Finally, we demonstrate the batch strategy for the cache blocking and the assignment of GPU computing resources such as threads and shared memory.
We also mention how to efficiently integrate our batched approaches to GCNs application.
In our supposed scenario of GCNs application, the format conversion for all sparse matrices causes expensive overhead. Thus, we focus on developing algorithms for simple sparse matrix formats such as COO and CSR.
Our target application is implemented with TensorFlow, and stores sparse matrix data as SparseTensor data structure.
We assume that the non-zero elements are not sorted by row or column indices in SparseTensor data structure.
The column size of dense input matrix in SpMM operation is the size of model in the application. Each column size of dense input in the batch is same.

\subsection{Sub-Warp-Assigned SpMM}
The SparseTensorDenseMatMul in TensorFlow launches $nnz_A * n_B$ threads, and one thread computes one multiplication-and-addition. Thus, $n_B$ threads are assigned to each non-zero element.
However, the actual CUDA implementation requires many operations for each thread to find which non-zero element or column is assigned, reducing the efficiency of parallelism.
Furthermore, when $n_B$ becomes larger, especially overcomes 32, more threads access to same non-zero element. This increases memory access demand and instruction counts. That is why the SparseTensorDenseMatMul results in lower performance.
We propose Sub-Warp-Assigned (SWA) SpMM, which assigns up to 32 threads (1 warp) to each non-zero element or row.
The $subWarp$ is set based on the column-size of dense input matrix, i.e. $n_B$.
\[
subWarp = \begin{cases}
  32 & (n_B > 16) \\
  \min{2^p} ~s.t.~ n_B \leq 2^p & (n_B \leq 16)
\end{cases}
\]
The number of assigned threads to each non-zero element is set as the power of two.
This enables to compute division and modulo operations by executing low-cost bit operations.
Furthermore, by setting the limit of assigning threads up to 32 threads to each non-zero element, the SWA SpMM reduces redundant memory pressure and instruction counts. 
We firstly show the algorithm of SWA SpMM for SparseTensor data structure, and then show that for CSR format.
The algorithm for SparseTensor is easily applied to COO format.

Figure~\ref{algo:code_swa_spmm_st} shows the pseudo CUDA code of SWA SpMM algorithm for SparseTensor data structure.
The SWA SpMM algorithm assigns $subWarp$ of threads to each non-zero element. The memory access to the dense input matrix and output matrix by threads among same $subWarp$ is coalesced.
As well as SparseTensorDenseMatMul, the SWA SpMM algorithm is load balanced since the work is parallelized by the number of non-zero elements.
Since another $subWarp$ assigned to different non-zero element has a chance of accessing same entry of output matrix, add operation is atomically executed.

Figure~\ref{algo:code_swa_spmm_csr} shows the pseudo CUDA code of SWA SpMM algorithm for CSR format.
By utilizing the characteristic of managing non-zero elements by row, the SWA SpMM for CSR assigns $subWarp$ to each row.
As well as SWA SpMM for SparseTensor, the threads among same $subWarp$ access to same non-zero element, and the memory access to dense input and output matrix is coalesced.
The significant difference from the SWA SpMM for SparseTensor is that the algorithm for CSR causes no data race on accessing output matrix. The product can be added without atomic operation.

\begin{figure}[t]
  \begin{center}
\begin{codebox}
    \Procname{$\procdecl{SWA\_SpMM\_ST}(C, A, B, subWarp)$}
\li \Comment $A$ is stored as SparseTensor data structure
\li \Comment set matrix $C$ to $O$
\li $\id{i} \gets \id{threadid}$
\li $\id{nzid} \gets \id{i} / subWarp$
\li $rid \gets ids_A[nzid * 2]$
\li $cid \gets ids_A[nzid * 2 + 1]$
\li $val \gets values_A[nzid]$
\li \For $\id{j} \gets (\id{i} \% subWarp)$ \To $n_B$ \By subWarp
\li \Do {\bf Atomic}($C[rid][j] \gets C[rid][j] + val * B[cid][j]$)
\End
\end{codebox}
  \end{center}
\vspace{10pt}
\caption{Pseudo CUDA code of Sub-Warp-Assigned SpMM for SparseTensor data structure}
\label{algo:code_swa_spmm_st}
\end{figure}

\begin{figure}[t]
  \begin{center}
\begin{codebox}
    \Procname{$\procdecl{SWA\_SpMM\_CSR}(C, A, B, subWarp)$}
\li \Comment $A$ is stored as CSR format
\li \Comment set matrix $C$ to $O$
\li $\id{i} \gets \id{threadid}$
\li $\id{rid} \gets \id{i} / subWarp$
\li \For $\id{nzid} \gets rpt_A[rid]$ \To $rpt_A[rid + 1]$
\li \Do $cid \gets colids_A[nzid]$
\li $val \gets values_A[nzid]$
\li \For $\id{j} \gets (\id{i} \% subWarp)$ \To $n_B$ \By subWarp
\li \Do $C[rid][j] \gets C[rid][j] + val * B[cid][j]$
\End
\End
\end{codebox}
  \end{center}
\vspace{10pt}
\caption{Pseudo CUDA code of Sub-Warp-Assigned SpMM for CSR format}
\label{algo:code_swa_spmm_csr}
\end{figure}

\subsection{Efficient use of shared memory and cache blocking}
Utilizing shared memory and improving the locality of memory access highly contribute to the performance of matrix multiplication on GPU.
Shared memory is implemented on each SM, and provides fast memory access. Also, hardware support improves the performance of atomic operations on shared memory.
In our approach for both SparseTensor data structure and CSR format, output matrix is temporally placed on shared memory.

\begin{figure*}[t]
\begin{center}
  \includegraphics[width=0.95\hsize]{\fig_dir/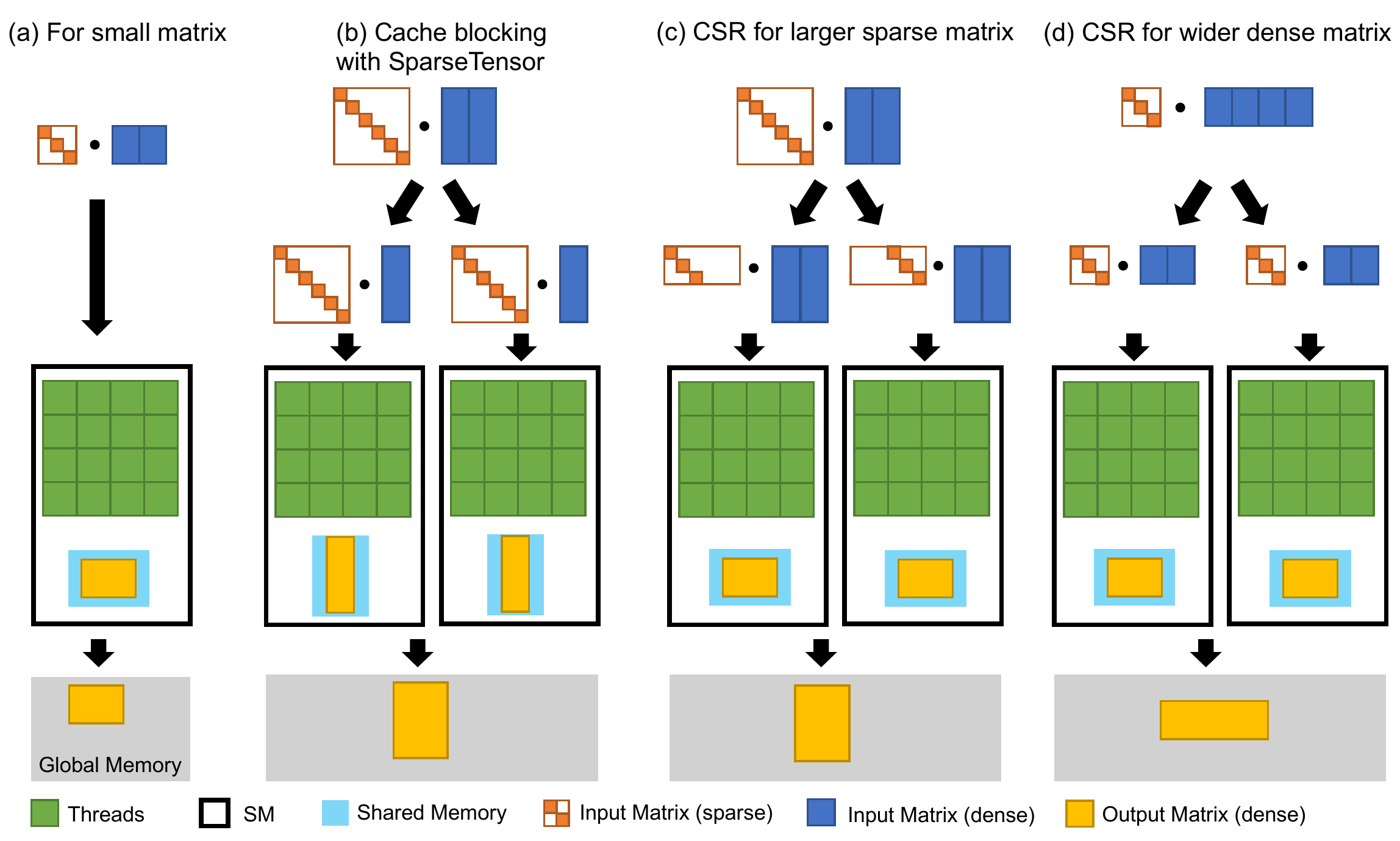}
  \caption{Utilization of shared memory and cache blocking optimization for SpMM}
  \label{fig:batched_spmm_algo}
\end{center}
\end{figure*}

For small matrix stored as COO format or SparseTensor data structure, our approach utilizes shared memory as Figure~\ref{fig:batched_spmm_algo}-(a). Utilizing shared memory for output matrix reduces the overhead of CUDA kernel launch for initializing output matrix.
As Figure~\ref{algo:code_swa_spmm_st} shows, the output matrix needs to be initialized with zero before actual SpMM operation starts.
The overhead of another CUDA kernel launch for initialization is not negligible, especially for SpMM on small matrices.
If $n_B$ or sparse matrix is large, whole output matrix can not be placed on shared memory of single streaming multiprocessor (SM).
In this case, cache blocking optimization divides the output matrix along the column, as Figure~\ref{fig:batched_spmm_algo}-(b) shows.
The cache blocking optimization enables SWA SpMM to utilize fast shared memory and achieve high performance even for larger input matrices.
Not only for output matrix, the locality of memory access to input dense matrix is also improved.

While each $subWarp$ in SWA SpMM for SparseTensor specifies only which column of output matrix is assigned without index information of non-zero element, the SWA SpMM for CSR can easily specify which row and column of output matrix is assigned.
Thus, shared memory does not need to keep whole output matrix. A $subWarp$ only needs shared memory with the size of $n_B$.
If $TB / subWarp * n_B$, $TB$ is thread block size, exceeds the capacity of shared memory, the cache blocking optimization is applied as Figure~\ref{fig:batched_spmm_algo}-(d) shows. As well as SparseTensor data structure, the output matrix and input dense matrix are divided along the column.

\subsection{Batched Algorithm for SpMM}
Our Batched SpMM manages tens or hundreds of SpMM operations by single CUDA kernel launch.
Based on the sizes of input sparse and dense matrices, the Batched SpMM algorithm decides whether the cache blocking is applied and how many threads are assigned to whole SpMM operations.

The Batched SpMM for SparseTensor decides whether cache blocking optimization is applied by the maximum size of output matrix in batch, i.e. $\max{m_A} * n_B$.
There are three cases by the sizes of input matrices.
\begin{enumerate}
\item Whole output matrix is small enough to be placed on shared memory (Figure~\ref{fig:batched_spmm_algo}-(a)) \label{enum:small}
\item Sub output matrix divided by cache blocking can be placed on shared memory (Figure~\ref{fig:batched_spmm_algo}-(b)) \label{enum:middle}
\item A graph (sparse matrix) is too large to use shared memory even with cache blocking optimization \label{enum:large}
\end{enumerate}
However, when we assume that the size of shared memory to each SpMM operation in single precision is 32KB, only the input sparse matrices with $m_A > 8192$ require the case~\ref{enum:large}). In this case, the SpMM operations do not need to be batched, and it is better option to execute single SpMM operation by optimized kernel for large input sparse matrix.
We focus on only the cases~\ref{enum:small}) and \ref{enum:middle}) although we support the case~\ref{enum:large}) as the implementation without using shared memory.
The Batched SpMM algorithm applies cache blocking optimization to all SpMM operations in the batch if an output matrix which cannot be placed on shared memory exists.
The Batched SpMM assigns one thread block to each SpMM operation for whole matrix or sub matrix.
For example, we consider that the Batched SpMM executes one hundred of SpMM operations in single precision and the capacity of shared memory assigned to each thread block is 32KB.
If any output matrix in the batch can be placed on shared memory without cache blocking, that is $m_A * n_B * \mathrm{sizeof(float)} \leq 32*1024$, the Batched SpMM launches one hundred thread blocks and each thread block computes one SpMM operation.
If one SpMM operation in the batch requires cache blocking and the output matrix is divided into two sub matrices, two hundreds thread blocks are launched and each thread block is assigned to one SpMM operation on sub matrix.

The required number of threads for each SpMM operation in the Batched SpMM for CSR format is simply $subWarp * m_A$.
As Figure~\ref{fig:batched_spmm_algo}-(c) shows, larger input matrix increases the number of thread blocks launched. 
The Batched SpMM for CSR format applies cache blocking optimization only when $n_B$ is large as Figure~\ref{fig:batched_spmm_algo}-(d) shows.
If the column size of dense input, $n_B$, is small enough, the Batched SpMM launches $\max{m_A} * subWarp * batch$ threads as a whole.
If wider input dense matrix needs cache blocking optimization and the output matrix is divided into $p$ sub matrices, the Batched SpMM launches $\max{m_A} * subWarp * batch * p$ threads as a whole.
Although the Batched SpMM for CSR launches more threads than necessary if the batch includes various sizes of input sparse matrices, these redundant threads do not become a problem since the threads terminate immediately.

\subsection{Batched SpMM for GCNs Application}
Our target GCNs application named ``ChemGCN'' constructs a neural network with multiple graph convolution layers.  The application is written with TensorFlow, but our batched approach does not limit the framework or implementation.
In our implementation with TensorFlow, we focus on applying the Batched SpMM for SparseTensor, but the Batched SpMM for CSR format can also be utilized in another framework such as Chainer, which supports CSR format.
Figure~\ref{algo:graph_conv} shows the flow of the graph convolution layer in ChemGCN.
The graph convolution layer executes convolutional operation for each input data. Firstly, the feature vectors on each vertex within the input graph are multiplied by the weight matrix for each channel (MatMul).
The bias is added to the multiplication result (Add), and the result is convoluted along with graph structure (SpMM).
Finally, the output of SpMM for each channel is accumulated.
This GraphConvolution layer requires $batchsize * channel$ times of kernel launches for MatMul, Add and SpMM. Since input matrices are small, the computing kernels for these three operations hardly exploit high parallelism of GPU. The overhead of CUDA kernel launches becomes critical performance issue.
The implementation insufficient for the performance is simply due to no availability of batched approach for SpMM.

Figure~\ref{algo:graph_conv_batched} shows the algorithm of graph convolution layer with the Batched SpMM.
To attain more parallelism, the operations within mini batch are batched.
The Batched SpMM applied to the graph convolution layer enables MatMul and Add to be also executed by single kernel, respectively.
Due to the limitation of the framework, the dimensions of input tensors are changed. In Figure~\ref{algo:graph_conv_batched}, the input dense tensor is reshaped from $\mathbb{R}^{m_X \times n_X \times batchsize}$ to $\mathbb{R}^{(m_X * batchsize) \times n_X}$.
However, this operation just changes the meta data of input data, and brings only a little overhead.
Generating the list of adjacency matrices for executing Batched SpMM requires only accumulating the pointers to the objects.
In this way, the operations on the graph convolution layer can be batched with little overhead.
The CUDA kernel launches for computing convolution are largely reduced from $O(channel * batchsize)$ to $O(channel)$.
The Batched SpMM is also applied to backward propagation.
\begin{figure}[t]
  \begin{center}
\begin{codebox}
    \Procname{$\procdecl{GraphConvolution}(Y, A, X, W, bias)$}
\li \For $b \gets 0$ \To $batchsize$
\li \Do \For $ch \gets 0$ \To $channel$
\li \Do $U \gets \procdecl{MatMul}(X[b], W[ch])$
\li $B \gets \procdecl{Add}(bias[ch], U)$
\li $C[ch] \gets \procdecl{SpMM}(A[b][ch], B)$
\End
\li $Y[b] \gets \procdecl{ElementWiseAdd}(C)$
\End
\end{codebox}
  \end{center}
\vspace{10pt}
\caption{Pseudo code of graph convolution layer in GCN application without batched optimization}
\label{algo:graph_conv}
\end{figure}

\begin{figure}[t]
  \begin{center}
\begin{codebox}
    \Procname{$\procdecl{GraphConvolutionBatched}(Y, A, X, W, bias)$}
\li \For $ch \gets 0$ \To $channel$
\li \Do $Xr \gets \procdecl{Reshape}(X, (m_X*batchsize, n_X))$
\li $U \gets \procdecl{MatMul}(Xr, W[ch])$
\li $B \gets \procdecl{Add}(bias[ch], U)$
\li $A_{list} \gets [A[0][ch], ..., A[batchsize - 1][ch]]$
\li $C[ch] \gets \procdecl{BatchedSpMM}(A_{list}, B)$
\End
\li $Y \gets \procdecl{ElementWiseAdd}(C)$
\end{codebox}
  \end{center}
\vspace{10pt}
\caption{Pseudo code of graph convolution layer in GCN application with batched optimization}
\label{algo:graph_conv_batched}
\end{figure}

\section{Performance Evaluation}
We evaluated the performance and effectiveness of the Batched SpMM. First we conducted the preliminary evaluation on randomly generated matrices to see the performance difference of SpMM between the non-batched and batched approaches. Next we evaluated the performance of ChemGCN application with the Batched SpMM.

We used TSUBAME3.0 at Tokyo Institute of Technology, which has two sockets of Intel(R) Xeon(R) CPU E5-2680 v4 @ 2.40GHz and four cards of NVIDIA Tesla P100-SMX2 GPU per computing node. In our evaluation, we used only one GPU. The Tesla P100 GPU consists of 56 SMs, which are 3584 of CUDA cores, and has 16GB HBM2 main memory with 732GB/sec peak memory bandwidth. The shared memory with 64KB is implemented on each SM, and 4MB L2 cache is shared among all threads. The OS was SUSE Linux Enterprise Server 12 (x86\_64), and the version of CUDA was 9.0.176. Our target GCNs application is implemented with TensorFlow. The version of TensorFlow was 1.8.0, and cuDNN was ver.7.0. All evaluation in this paper was in single precision.

\subsection{Preliminary Evaluation Results}
We evaluated the performance of batched approaches on randomly generated sparse matrix data.
We used SpMM functions from cuSPARSE library with CSR format. We evaluated both ``csrmm()'' and ``csrmm2()'', and show the best one as ``cuSPARSE''. We implemented SpMM function with SparseTensor-like format following SparseTensorDenseMatMul as the non-batched approach. Compared to the non-batched approaches, we evaluated Batched SpMM for CSR and SparseTensor-like formats. In addition to them, we evaluated the performance of ``gemmBatched()'' from cuBLAS library by treating the input sparse matrix as dense matrix.
This Batched GEMM function is optimized for processing many GEMM operations by single kernel.
The Batched GEMM can exploit GPU's computing power while most of operations are zero-related.
Although we just show the throughput of Batched GEMM as evaluation result, holding sparse matrix as dense format requires much memory usage.
When we consider the practical application scenario, much more memory requirement by dense format makes it hard to keep the data on GPU memory, causing more memory transfer between CPU and GPU.
Since our target is graph data, the randomly generated sparse matrices are square. The row size ($dim$) and $nnz/row$ are parameterized in generating matrix, and the non-zero pattern is different from each other. The column size of input dense matrix ($n_B$) is also parameterized.
The performance number is showed with FLOPS, calculated by $2 * nnz_A * n_B / exe\_time$.
Since the concern is actual execution time, the ``gemmBatched()'' also follows this performance metric although many zero-related operations are executed.
The Batched SpMM and Batched GEMM require the pointer to each matrix data to be placed on device memory, while it is stored on host memory in the non-batched cases. Our evaluation for batched approaches includes memory transfer of pointer arrays from host to device.
Each performance number is the average of 10 times executions.

\begin{figure*}[t]
\subfloat[][$batchsize=50, dim=50, nnz/row=2$]
{
  \includegraphics[width=0.5\hsize]{\fig_dir/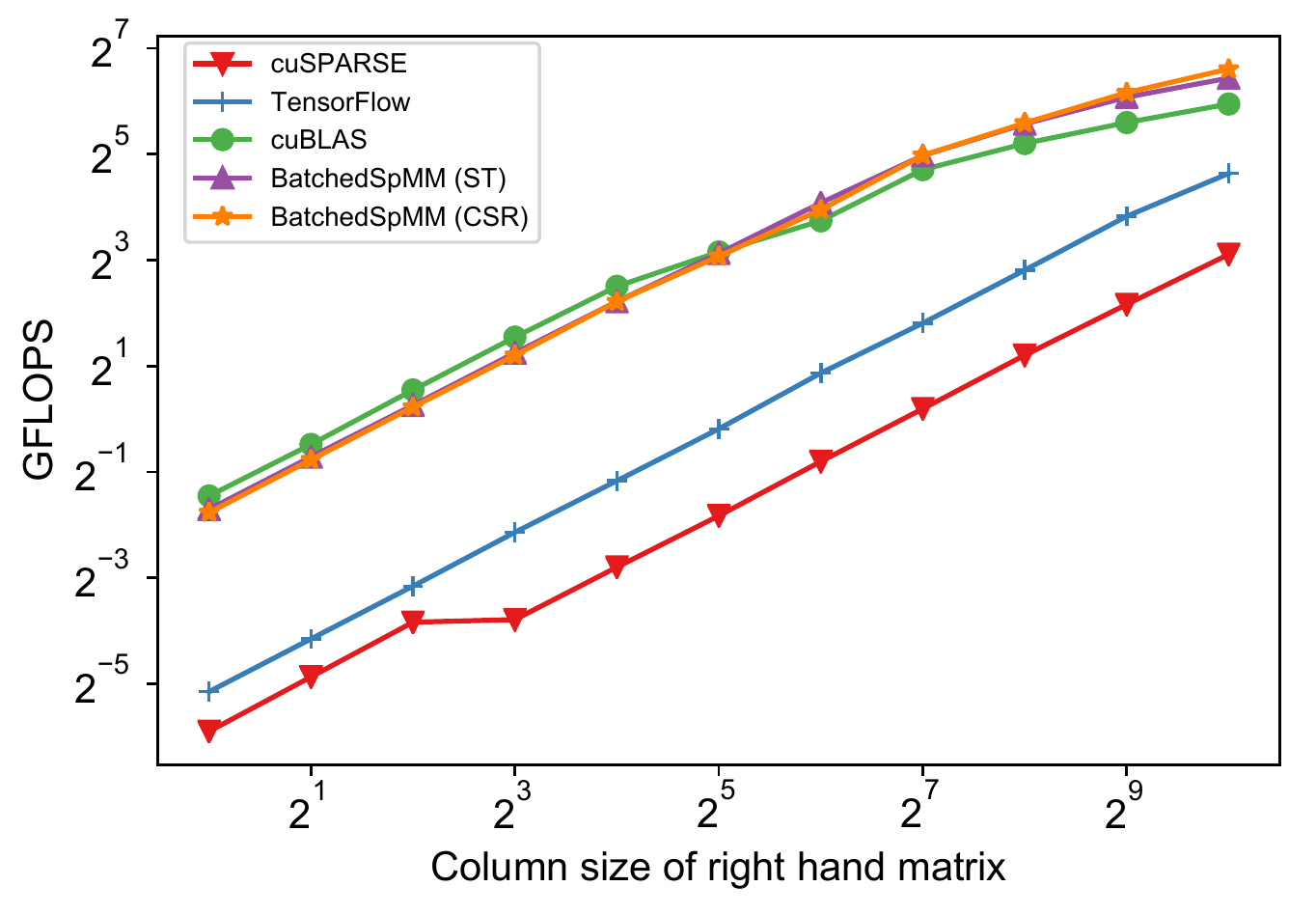}
 }
~
\subfloat[][$batchsize=100, dim=50, nnz/row=3$]
{
  \includegraphics[width=0.5\hsize]{\fig_dir/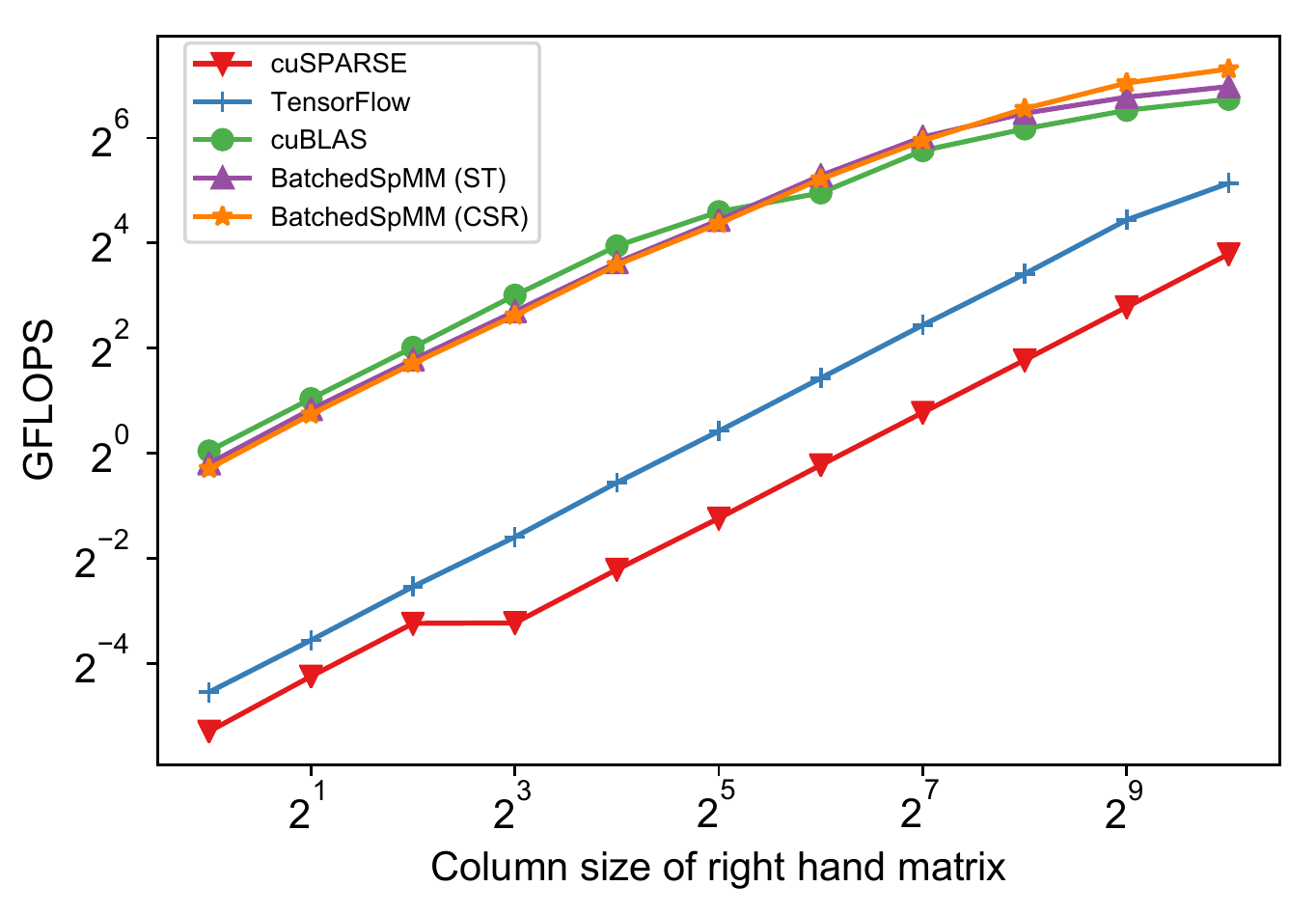}
}
  \caption{Performance of SpMM and Batched GEMM on randomly generated dataset following the one used in target GCNs application. BatchedSpMM (ST) represents the result of Batched SpMM for Sparse Tensor.}
  \label{fig:bbench_all}
\end{figure*}

Figure~\ref{fig:bbench_all} shows the performance of non-batched and batched approaches for SpMM.
The parameter setting for the input sparse matrices in the evaluation is based on dataset and configurations of GCNs application showed in later part of this section. The evaluation result in Figure~\ref{fig:bbench_all} works as a proxy for estimating the impact of Batched SpMM applied to the GCNs application.
In both cases in Figure~\ref{fig:bbench_all}, the Batched SpMM shows best performance on fat dense matrices.
The non-batched approaches, which sequentially process SpMM, cannot achieve high performance due to the overhead of repeated CUDA kernel launches and poor parallelism of GPU. On the other hand, all of the batched approaches acquire the large performance gain.
Compared to the implementation of SpMM following TensorFlow, the Batched SpMM achieves significant speedups of up to 9.27x at $n_B=64$ in Figure~\ref{fig:bbench_all}-(a) and 6.09x at $n_B=512$ in Figure~\ref{fig:bbench_all}-(b).
\New{To dive into more detailed analysis, we evaluated the {\it sm\_efficiency} by using nvprof, which is one of the profiler tools for NVIDIA GPU.
The {\it sm\_efficiency} is the percentage of active streaming multiprocessors (SMs).
While the {\it sm\_efficiency} with non-batched implementation of SpMM following TensorFlow is 35.51\% at $n_B=512$ in Figure~\ref{fig:bbench_all}-(b), the Batched SpMM for SparseTensor and for CSR show 89.07\% and 87.87\%, respectively. 
This result simply shows that the Batched SpMM exploits GPU's computing resources.}

The Batched GEMM routine of cuBLAS also shows high throughput although it includes many zero-related operations.
This is because the target sparse matrices are small and do not include so much zero-related operations.
It should be noted, however, that treating as sparse matrix and executing as SpMM is natural, and the Batched SpMM overcomes cuBLAS. The performance improvement of Batched SpMM from cuBLAS is 1.26x at $n_B=64$ in Figure~\ref{fig:bbench_all}-(a) and 1.43x at $n_B=512$ in Figure~\ref{fig:bbench_all}-(b).
In the cases with smaller $n_B$, the Batched GEMM of cuBLAS shows superior performance to our Batched SpMM. This is because the actual execution time on GPU is too short and the overhead for more memory transfer from CPU to GPU with Batched SpMM compared to Batched GEMM becomes dominant.

\begin{figure*}[t]
\subfloat[][$batchsize=100, dim=32, nnz/row=3$]
{
  \includegraphics[width=0.31\hsize]{\fig_dir/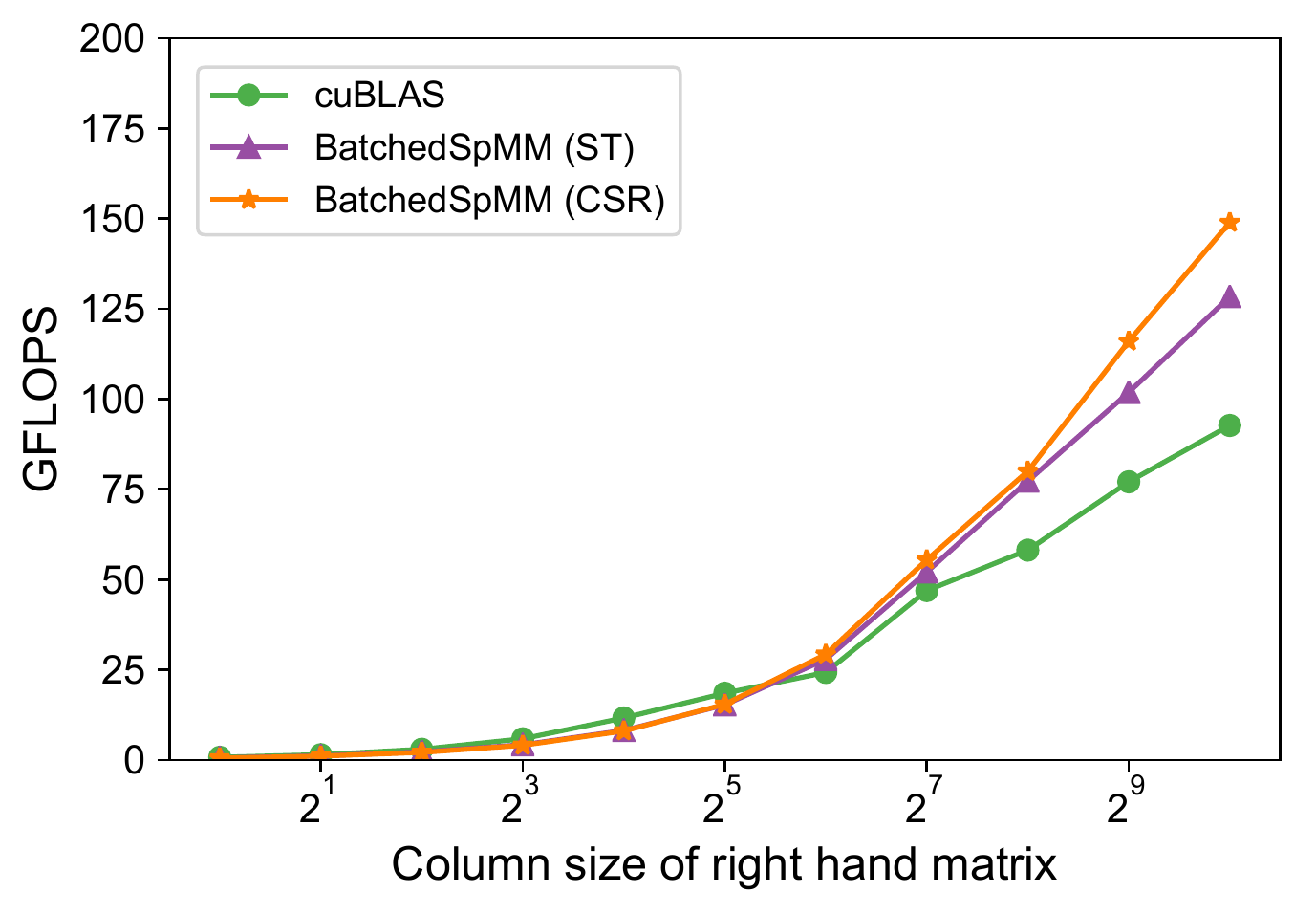}
}
~
\subfloat[][$batchsize=100, dim=64, nnz/row=3$]
{
  \includegraphics[width=0.31\hsize]{\fig_dir/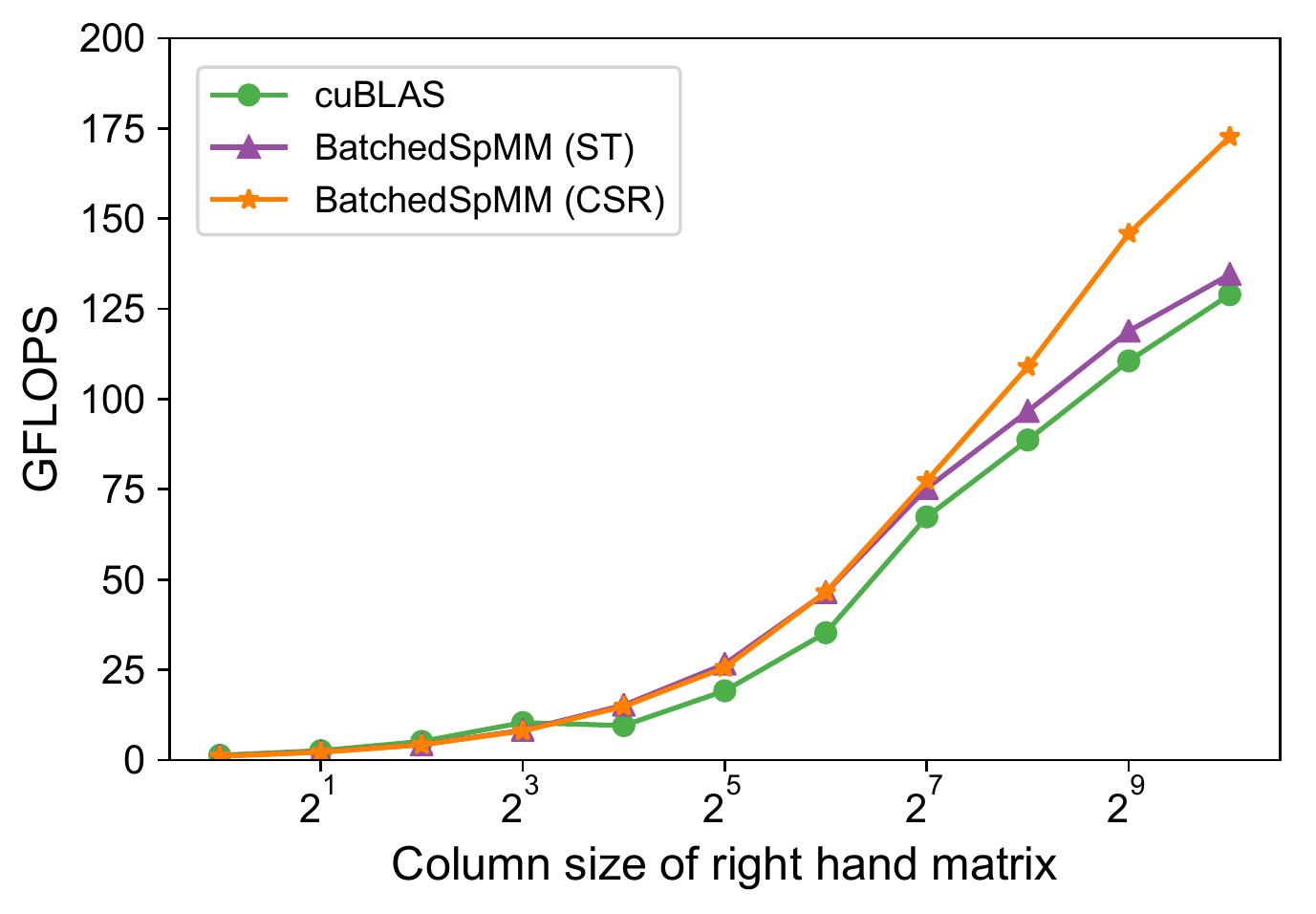}
}
~
\subfloat[][$batchsize=100, dim=128, nnz/row=3$]
{
  \includegraphics[width=0.31\hsize]{\fig_dir/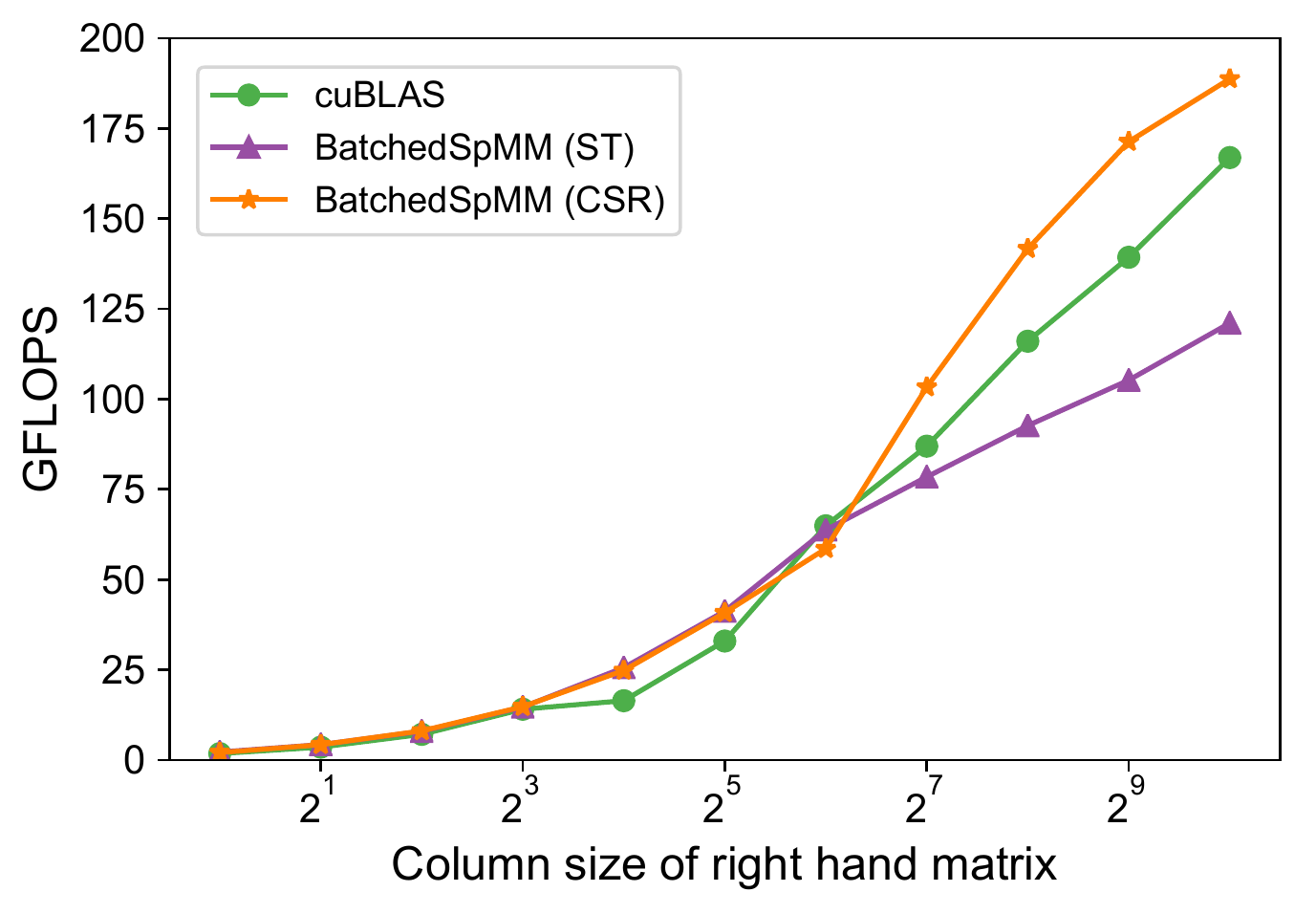}
}
\\
\subfloat[][$batchsize=50, dim=64, nnz/row=3$]
{
  \includegraphics[width=0.31\hsize]{\fig_dir/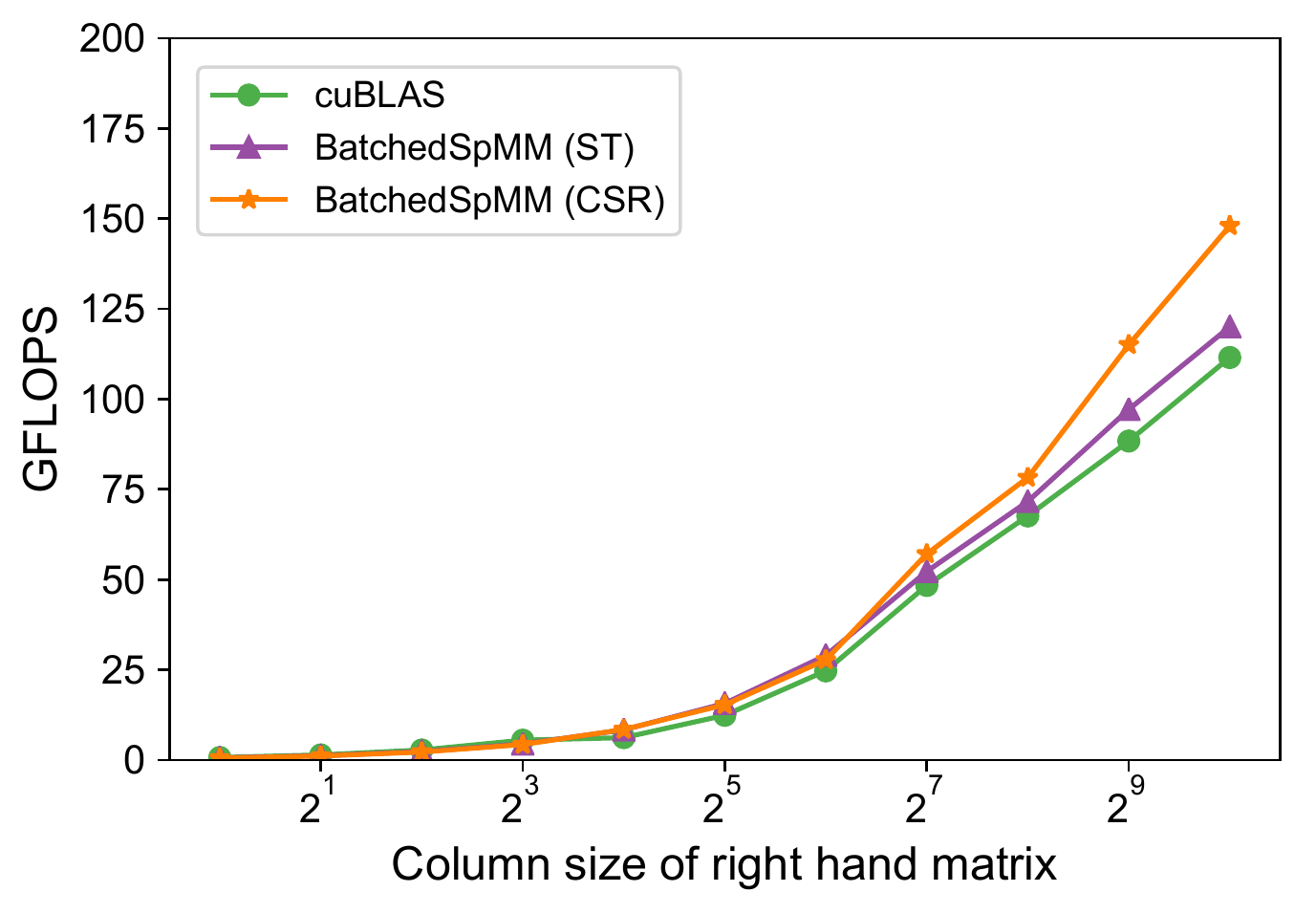}
}
~
\subfloat[][$batchsize=100, dim=64, nnz/row=1$]
{
  \includegraphics[width=0.31\hsize]{\fig_dir/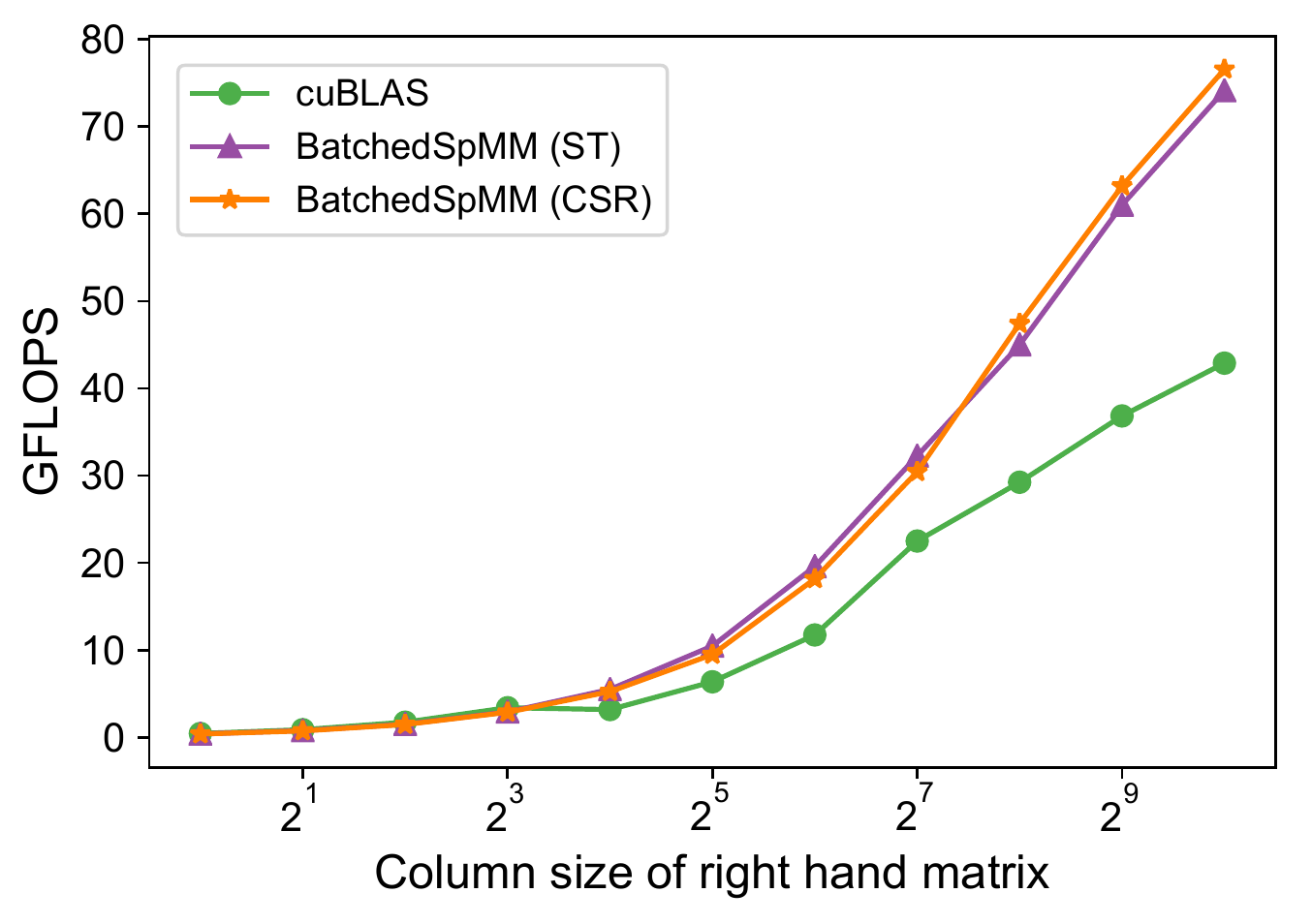}
}
~
\subfloat[][$batchsize=100, dim=64, nnz/row=5$]
{
  \includegraphics[width=0.31\hsize]{\fig_dir/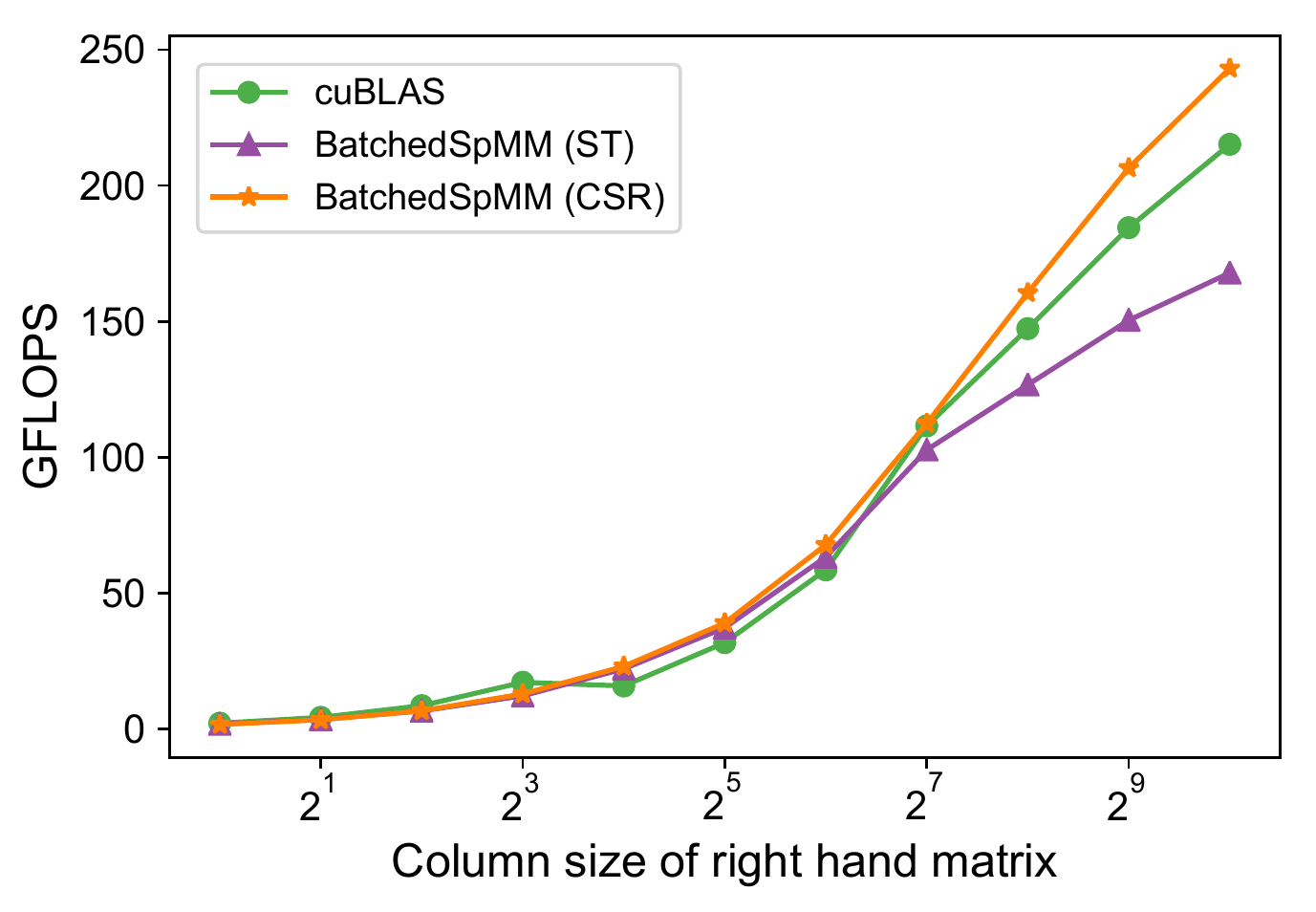}
}
  \caption{Performance of batched approaches of SpMM on randomly generated dataset}
  \label{fig:bbench}
\end{figure*}

Next, we show the evaluation results of three batched approaches with changing the parameters of input matrices and batch size.
The Figure~\ref{fig:bbench}-(b) and (d) show the performance difference between $batchsize$.
The results indicate that more batch size brings more throughput to all batched approaches.
The batched approaches with $batchsize=50$ fail in using all SMs on GPU. While small batch size hardly exploits the computing power of GPU, larger batch size, $batchsize=100$, ensures enough parallelism.
The first row of Figure~\ref{fig:bbench} (i.e. (a), (b) and (c)) shows the result of $dim=32, 64$ and $128$, respectively.
By increasing the size of input sparse matrix, the Batched SpMM for CSR and cuBLAS are getting better performance numbers.
This performance boost simply comes from the improvement of parallelism.
The Batched SpMM for CSR format launches more threads in proportion to the row size of input sparse matrix.
Thus, a larger input sparse matrix can improve the occupancy of GPU.
While the cuBLAS also increases the parallelism on larger input sparse matrix, the performance improvement is relatively smaller than Batched SpMM for CSR since the sparsity is also increased.
The Batched SpMM for SparseTensor shows only slight performance change.
This is because more cache blocking in Batched SpMM for SparseTensor causes more memory pressure to same non-zero element.

The Figure~\ref{fig:bbench}-(e) and (f) show the evaluation results of $nnz/row=1$ and $5$, respectively.
Obviously, the Batched SpMM works efficiently on sparser matrices, and the cuBLAS appears to show better performance on denser matrices.
For low nnz/row matrices, the Batched SpMM both for SparseTensor and for CSR are superior to cuBLAS.
On the other hand, the performance improvement by Batched SpMM for SparseTensor is limited on denser input sparse matrix due to more race condition by atomic operations.
The Batched SpMM for CSR is atomic-free algorithm, and keeps best performer on denser input sparse matrices.

\begin{figure}[t]
 \begin{center}
  \includegraphics[width=\hsize]{\fig_dir/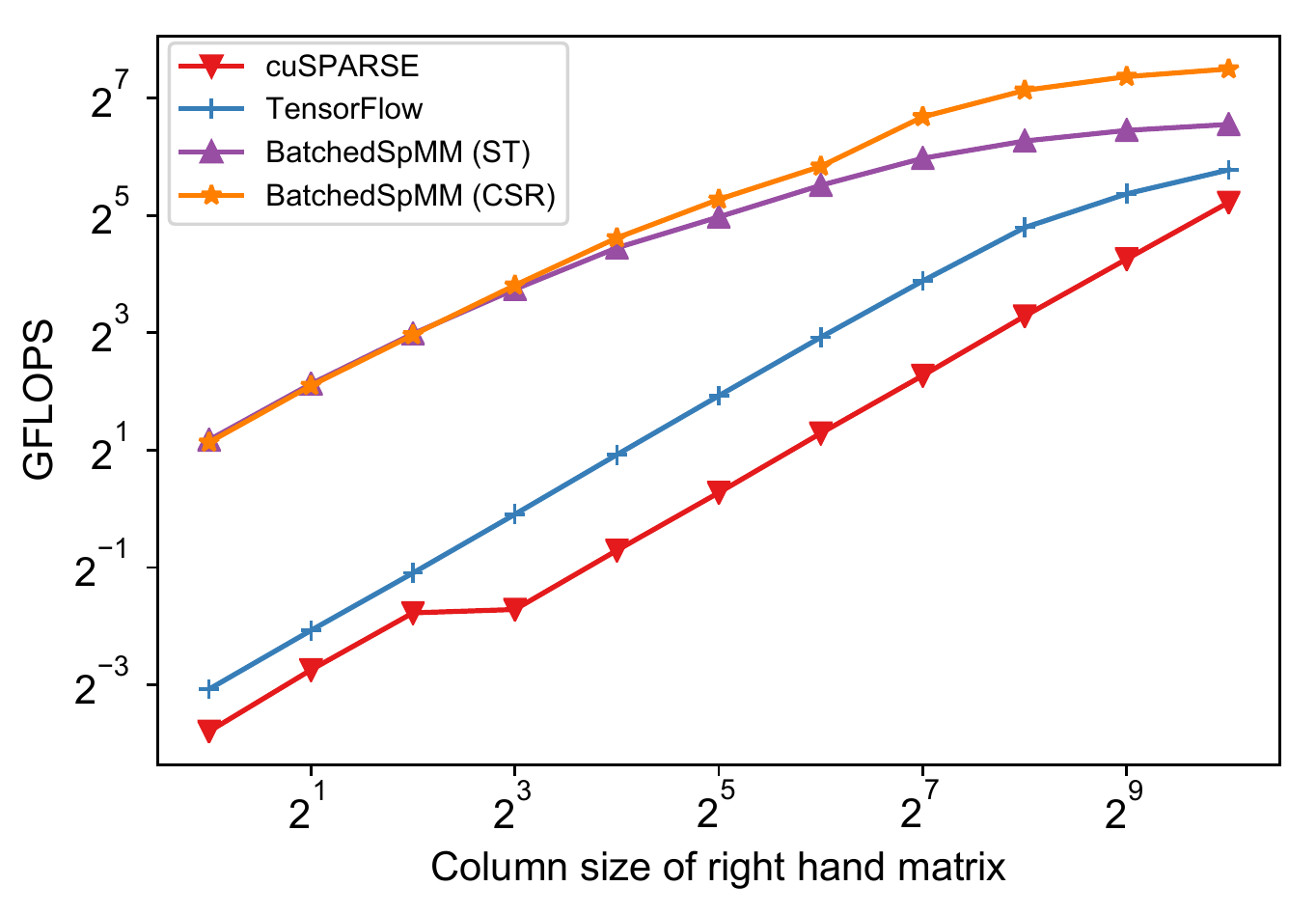}
  \caption{Performance of SpMM on randomly generated dataset ($batchsize=100$, $dim$ and $nnz/row$ are mixed.)}
  \label{fig:bbench_mix_100}
 \end{center}
\end{figure}

Finally, we consider the batch includes various sizes or densities of input sparse matrices.
Figure~\ref{fig:bbench_mix_100} shows the result with mixed sizes and densities, $batchsize=100, dim=[32, 256], nnz/row=[1,5]$.
We excluded cuBLAS from this evaluation since the kernel only processes GEMM operations with same matrix sizes.
The Batched SpMM achieves significant improvements compared to the non-batched approaches even though the matrices among batch have different shapes to each.
At $n_B=1024$, our Batched SpMM achieves up to 3.29x speedup from the non-batched approaches.

\begin{table}[t]
\small
\begin{center}
\caption{Dataset and configurations in the evaluation of ChemGCN application}
\label{tab:config}
\begin{tabular}{| l | r | r | r | r |}
\hline
& \#Matrices & Max $dim$ & Epoch & Batch size \\
\hline
Tox21 & 7,862 & 50 & 50 & 50 \\
Reaction100 & 75,477 & 50 & 20 & 100 \\
\hline
\end{tabular}
\end{center}
\end{table}

\subsection{Evaluation on GCNs Application}
We evaluated the impact of the Batched SpMM in ChemGCN application implemented with TensorFlow.
A graph convolutional network consists of several graph convolution layers. After each graph convolution layer, batch normalization is executed.
Table~\ref{tab:config} shows the dataset and the configurations for the application.
We used Tox21~\cite{tox21} and Reaction100 datasets, which predicts the reaction from its own graph structure of compound structure by selecting 100 kinds of typical reactions from Reaxys database\cite{reaxys}.

The \#Matrices in Table~\ref{tab:config} is the number of pairs of adjacency matrix and feature matrix.
The model is trained by K-fold cross validation, and k=5 in our evaluation. The trained model is used for the evaluation of inference.
The evaluation result of inference is the execution time for inferring all data of dataset.
In the evaluation of inference, the batch size is set as 200 to increase the throughput since the batch size does not affect the accuracy of inference.
The evaluation results of both training and inference are the mean of 5 executions. We used the Batched SpMM for SparseTensor data structure as the batched approach.
It should be noted that our batched optimization does not change the configuration or hyper parameters from non-batched version, and thus no effect on the accuracy in training.
The architecture for Tox21 includes two graph convolution layers. The column size of each weight matrix is 64.
For Reaction100, three graph convolutional layers are stacked, and the column size of each weight matrix is 512.

\begin{table}[t]
\begin{center}
\caption{Training time of ChemGCN [sec]}
\label{tab:gcnn_result_training}
\begin{tabular}{| l | r | r | r | r |}
\hline
 & CPU & \multicolumn{2}{|c|}{GPU} & \multirow{2}{*}{\bf Speedup} \\
\cline{2-4}
 & Non-Batched & Non-Batched & Batched & \\
\hline
Tox21 & 854.51 & 918.03 & 723.80 & {\bf 1.18x} \\
Reaction100 & 16223.98 & 3029.13 & 1905.32 & {\bf 1.59x} \\
\hline
\end{tabular}
\end{center}
\end{table}

\begin{table}[t]
\begin{center}
\caption{Inference time of ChemGCN [sec]}
\label{tab:gcnn_result_inference}
\begin{tabular}{| l | r | r | r | r |}
\hline
 & CPU & \multicolumn{2}{|c|}{GPU} & \multirow{2}{*}{\bf Speedup} \\
\cline{2-4}
 & Non-Batched & Non-Batched & Batched & \\
\hline
Tox21 & 2.71 & 2.56 & 1.97 & {\bf 1.30x} \\
Reaction100 & 44.66 & 22.42 & 16.32 & {\bf 1.37x} \\
\hline
\end{tabular}
\end{center}
\end{table}

Table~\ref{tab:gcnn_result_training} and \ref{tab:gcnn_result_inference} show the performance of training and inference in ChemGCN application, respectively. The training of Tox21 dataset on CPU finishes in less execution time compared to that on GPU with non-batched approach. This is because Tox21 dataset is small enough that all data such as adjacency matrices and feature matrices can be placed on last-level cache of CPU. The non-batched approach does not exploit the computing power of GPU due to the lack of parallelism, and the overhead of repeated CUDA kernel launches degrades the performance.
In contrast, the batched approach on GPU significantly improves the performance of GCNs application. Especially on Reaction100 dataset, which has more data and adopts larger batch size, the benefit of batched approach is prominent and the speedup of training compared to the non-batched approaches achieves 1.59x. We can maximize the benefit of the batched approach in the inference by setting larger batch size. The batched approach achieves up to 1.39x speedup of inference throughput.

\begin{figure*}[t]
 \begin{center}
  \includegraphics[width=\hsize]{\fig_dir/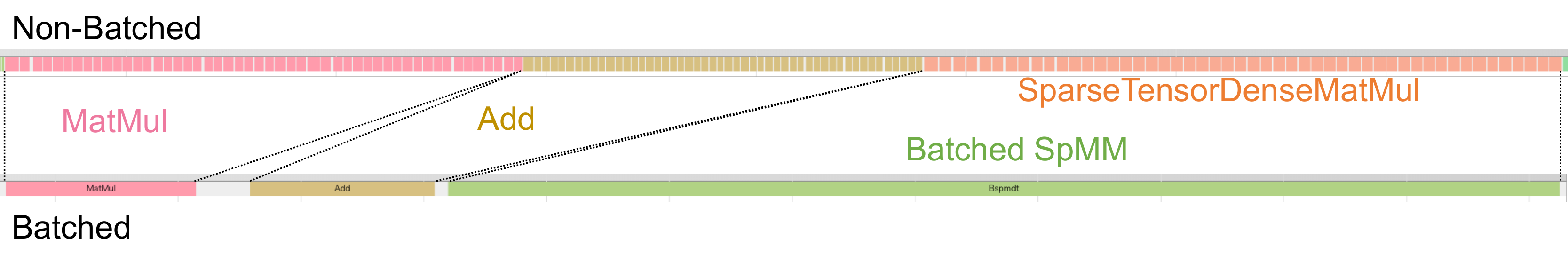}
  \caption{Visualization of the execution of ChemGCN with Timeline}
  \label{fig:timeline}
 \end{center}
\end{figure*}

Next, we analyzed the performance of each kernel by using Timeline, which is a performance profiling tool in TensorFlow. Figure~\ref{fig:timeline} shows the visualization of each kernel execution in a graph convolutional layer for one mini batch on Tox21 dataset with the non-batched or batched approaches. Although the bars are illustrated with same length, their execution time are completely different. Table~\ref{tab:timeline} shows the ``actual'' execution time of each operation.
As Figure~\ref{fig:timeline} shows, the non-batched approach requires $batchsize*3=150$ times of CUDA kernel launches while the batched approach requires only three times. The batched approach can largely reduce the overhead of CUDA kernel launches. The speedup of SpMM with Batched SpMM is about 10x. This performance improvement corresponds to the number showed in Figure~\ref{fig:bbench_all}-(a).
Also, the batched approach significantly reduces the execution time of MatMul and Add as well as SpMM. This implies that the non-batched approach does not exploit the parallelism of GPU. We concluded that the batched approach can significantly increase the occupancy of GPU not only in SpMM operation but also in other kernels on small matrices.

\begin{table}[t]
\begin{center}
\caption{Execution time of each kernel in Figure~\ref{fig:timeline} [$\mu$sec]}
\label{tab:timeline}
\begin{tabular}{| l | r | r|}
\hline
 & Non-Batched & Batched \\
\hline
MatMul & 1,571 & 31 \\
Add & 1,316 & 23 \\
SpMM & 1,981 & 190 \\
\hline
\end{tabular}
\end{center}
\end{table}

\section{Conclusion}
In the application of machine learning approaches to various fields including chemistry and biology, GCNs show high accuracy. However, there are yet less work of GCNs in terms of throughput, and the current approach for GCNs, which require many operations on small sparse and dense matrices, cannot exploit the computing power of GPU. 
The scenario requiring small sparse matrix operations, especially SpMM, does not yet exist, and thus no research or engineering effort has been put into accelerating for such kind of operations.
In order to accelerate GCNs applications, we propose the Batched SpMM, which efficiently handles tens or hundreds of SpMM operations together by single kernel. To exploit GPU architecture more, we devise new SpMM algorithm, Sub-Warp-Assigned SpMM, especially for small sparse matrices with cache blocking optimization to dense matrices.
We evaluate the performance of our Batched SpMM on randomly generated matrices. The Batched SpMM achieves up to 9.27x speedup compared to the non-batched approaches, and 1.43x compared to Batched GEMM of cuBLAS. Furthermore, the batched optimization improves the performance of GCNs application with 1.59x speedup for training and 1.37x speedup for inference.

For future work, we are planning to have more analysis and optimization to improve the GCNs application since the hotspot of the application moves to other while our batched approach significantly improves the performance of SpMM-related operations.
Our batched approach contributes not only to the reduction of the overhead of kernel launches, but also to increasing the occupancy of computing devices. We are planning to evaluate whether our batched approach works well on other architectures such as multi-core CPUs or Intel KNL.

\section*{Acknowledgment}
This work was partially supported by JST CREST Grant Number JPMJCR1303 and JPMJCR1687, 
and performed under the collaboration with DENSO IT Laboratory, inc., 
and performed under the auspices of Real-World Big-Data Computation Open Innovation Laboratory, Japan, 
and based on results obtained from a project commissioned by the New Energy and Industrial Technology Development Organization (NEDO).
We would like to thank Mr. Akira Naruse of NVIDIA for providing much of advices on our research.

\bibliographystyle{IEEEtran}

\end{document}